\documentclass[twocolumn,showpacs,superscriptaddress,prb,floatfix]{revtex4}%
\usepackage{amsfonts}
\usepackage{amsmath}
\usepackage{amssymb}
\usepackage{graphicx}%
\providecommand{\U}[1]{\protect\rule{.1in}{.1in}}

\begin{document}
\title{Mechanical and chemical bonding properties of ground state BeH$_{2}$}
\author{Bao-Tian Wang}
\affiliation{Institute of Theoretical Physics and Department of
Physics, Shanxi University, Taiyuan 030006, People's Republic of
China} \affiliation{LCP, Institute of Applied Physics and
Computational Mathematics, Beijing 100088, People's Republic of
China}
\author{Ping Zhang}
\thanks{To whom correspondence should be addressed. Electronic address:
zhang\_ping@iapcm.ac.cn} \affiliation{LCP, Institute of Applied
Physics and Computational Mathematics, Beijing 100088, People's
Republic of China}
\author{Hongliang Shi}
\affiliation{SKLSM, Institute of Semiconductors, Chinese Academy of
Sciences, People's Republic of China} \affiliation{LCP, Institute of
Applied Physics and Computational Mathematics, Beijing 100088,
People's Republic of China}
\author{Bo Sun}
\affiliation{LCP, Institute of Applied Physics and Computational Mathematics, Beijing
100088, People's Republic of China}
\author{Weidong Li}
\affiliation{Institute of Theoretical Physics and Department of
Physics, Shanxi University, Taiyuan 030006, People's Republic of
China} 

\begin{abstract}
The crystal structure, mechanical properties and electronic structure of
ground state BeH$_{2}$ are calculated employing the first-principles methods
based on the density functional theory. Our calculated structural parameters
at equilibrium volume are well consistent with experimental results. Elastic
constants, which well obey the mechanical stability criteria, are firstly
theoretically acquired. The bulk modulus \emph{B}, Shear modulus \emph{G},
Young's modulus \emph{E} and Poisson's ratio $\upsilon$ are deduced from the
elastic constants. The bonding nature in BeH$_{2}$ is fully interpreted by
combining characteristics in band structure, density of state, and charge
distribution. The ionicity in the Be$-$H bond is mainly featured by charge
transfer from Be 2\emph{s} to H 1\emph{s} atomic orbitals while its covalency
is dominated by the hybridization of H 1\emph{s} and Be 2\emph{p} states. The
valency in BeH$_{2}$ can be represented as Be$^{1.99+}$H$^{0.63-}$, which
suggests that significant charge transfer process exists.

\end{abstract}
\maketitle

\section{INTRODUCTION}

There has been a great interest in simple alkali-metal and
alkaline-earth-metal hydride systems stimulated by both the fundamental
interests and various applications. Such kinds of metal hydrides, e.g.,
beryllium and magnesium and beryllium-magnesium-based hydrides, which all have
high weight percentage of hydrogen, are promising candidates for hydrogen
storage. If they become metallic when subjected to high pressures, they may be
candidates for high-temperature superconductivity. Among them, MgH$_{2}$ has
been extensively studied both experimentally and theoretically. Due to the
experimental difficulties in material synthesis \cite{Armstrong}, on the other
hand, a comprehensive understanding of BeH$_{2}$ is still being in progress.
It had long been considered that BeH$_{2}$ was a polymer and did not
crystallize without ternary additions. Recently, however, Smith \emph{et al}.
\cite{Smith} have successfully synthesized crystalline BeH$_{2}$ and first
established the crystal structure as the body-centered orthorhombic by
synchrotron-radiation-based powder x-ray diffraction. Since this experimental
establishment, to date theoretical studies on physical properties of
crystalline BeH$_{2}$ are very scarce in the literature. Vajeeston \emph{et
al}. \cite{Vajeeston} and Hantsch \emph{et al}. \cite{Hantsch} have studied
the structural stability of BeH$_{2}$ by performing first-principles
total-energy calculations for various types of structural variants, among
which they confirmed that the experimentally observed BeH$_{2}$ modification
has the lowest total energy. No further theoretical studies have been
reported, although more experimental works are going on to investigate the
pressure dependence of physical properties of BeH$_{2}$ \cite{Ahart}.

In this paper, we have carried out the first-principles calculations of
ground-state behavior of BeH$_{2}$ at its experimentally established
crystalline phase. Results for the electronic and atomic structure, mechanical
properties, charge density participation, and the nature of Be-H chemical
bonding, are systematically presented. Our calculated results for elastic
constants indicate that the experimentally observed orthorhombic phase of
BeH$_{2}$ is mechanically stable. We determined that the ionic charge of Be
and H in BeH$_{2}$ are represented Be$^{1.99+}$ and H$^{0.63-}$, which, when
compared to Mg$^{1.95+}$ and H$^{0.6-}$ in MgH$_{2}$, indicates that the
charge transfer process in BeH$_{2}$ is more prominent in MgH$_{2}$.

\section{Calculation method}

The density-functional theory (DFT) total energy calculations for the ground
state BeH$_{2}$ were carried out using the Vienna \textit{ab initio}
simulations package(VASP) \cite{Kresse3} with the projected-augmented-wave
(PAW) pseudopotentials \cite{PAW} and plane waves. The generalized gradient
approximation (GGA) \cite{GGA} for the exchange-correlation potential was
employed. 7$\times$7$\times$7 Monkhorst-Pack \cite{Monk} \emph{k} points in
the full wedge of the Brillouin zone were used. The plane-wave energy cutoff
was set 500 eV, and all atoms were fully relaxed until the Hellmann-Feynman
forces were less than 0.002 eV/\AA . To obtain optimized lattice vectors and
atomic coordinates of bulk unit cell, we relaxed it at a series of fixed
volumes. The bulk modulus \emph{B}, shear modulus \emph{G}, Young's modulus
\emph{E}, Poisson's ratio $\upsilon$, were gained through computing elastic constants.

\section{RESULTS AND DISCUSSION}

\subsection{Atomic structure and mechanical properties}

\begin{figure}[tbp]
\begin{center}
\includegraphics[width=0.6\linewidth]{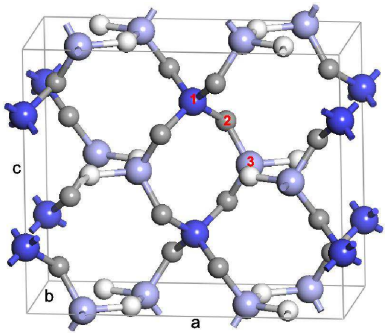}
\end{center}
\caption{(Color online) Unit cell for body-centered orthorhombic
BeH$_{2}$. Both Be and H have two kinds of atomic occupations, which
are in the figure presented by blue (Be1), light blue (Be2), gray
(H1), and white (H2) spheres.} \label{fig1}
\end{figure}%

\begin{table*}
\caption{Calculated structural parameters, bulk modulus \emph{B},
shear modulus \emph{G}, Young's modulus \emph{E}, Poisson's ratio
$\upsilon$ and energy band gap (\emph{E$_{g}$}) for orthorhombic
BeH$_{2}$. As a comparison,
other theoretical work and available experimental data are listed.}%
\label{tab:table1}
\begin{ruledtabular}
\begin{tabular}{ccccccc}
&Present calculations&Previous calculations&Expt.\\
\hline
\emph{a} ({\AA})& 9.0049 & 8.9823 & 9.082 \\
\emph{b} ({\AA})& 4.1684 & 4.1563 & 4.160 \\
\emph{c} (${\rm{\AA}}$)& 7.6347 & 7.6455 & 7.707 \\
\emph{V} ($${\AA}$^{3}$)& 286.6 & 285.4 & 291.2 \\
Coordinates & Be1(4\emph{a}): 0, 0, 0.25&0, 0, 0.25  & 0, 0, 0.25 \\
& Be2(8\emph{j}): 0.1684, 0.1199, 0 & 0.1678, 0.1207, 0 &0.1699, 0.1253, 0  \\
& H1(16\emph{k}): 0.0881, 0.2216, 0.1516& 0.0720, 0.1842, 0.1361 & 0.0895, 0.1949, 0.1515 \\
& H2(8\emph{j}): 0.3092, 0.2792, 0& 0.3097, 0.2790, 0 & 0.3055, 0.2823, 0 \\
Be1-H1 ({\AA})&1.328  &1.435$^{\emph{b}}$  &1.38(2) \\
Be2-H1 ({\AA})&1.376  &  &1.41(2) \\
Be2-H2 ({\AA})&1.434  &1.439$^{\emph{b}}$  &1.44(2) \\
H1-Be1-H1 ($^{\circ}$)&109.6  &  &107.7(5) \\
H1-Be2-H1 ($^{\circ}$)&98.3  &98.6$^{\emph{b}}$  &112.1(5) \\
H1-Be2-H2 ($^{\circ}$)&117.9  &  &111.2(5) \\
H2-Be2-H2 ($^{\circ}$)&109.2  &  &109.1(5) \\
Be1-H1-Be2 ($^{\circ}$)&133.6  &126.4$^{\emph{b}}$  &130(1) \\
Be2-H2-Be2 ($^{\circ}$)&125.4  &120.0$^{\emph{b}}$  &127(1) \\
\emph{B} (GPa)&22.60 &23.79 &14.2$\pm$3.0$^{\emph{a}}$ \\
\emph{G} (Gpa)&24.50  &  &  \\
\emph{E} (GPa)&53.99  &  & \\
$\upsilon$ & 0.102 &  & 0.21$^{\emph{a}}$ \\
\emph{E$_{g}$} (eV)&5.52&5.51&\\
\end{tabular}
\end{ruledtabular}
$^{\emph{a}}$Experimentally measured values from Ref. \cite{Ahart}
for amorphous BeH$_{2}$ at ambient pressure. Other experimental
values are selected from Ref. \cite{Smith}. $^{\emph{b}}$Previous
theoretical values from Ref. \cite{Hantsch} and other from Ref.
\cite{Vajeeston}.
\end{table*}

The ground state of BeH$_{2}$, which has been testified both
experimentally \cite{Smith} and theoretically \cite{Vajeeston},
belongs to body-centered orthorhombic structure with space group
\emph{Ibam}. Its orthorhombic unit cell is composed of twelve
BeH$_{2}$ formula units, see Fig. 1. The primitive cell, containing
six formula units, has two Be1 (4\emph{a}, in Wyckoff notation),
four Be2 (8\emph{j}), eight H1 (16\emph{k}), and four H2 (8\emph{j})
atoms. Each Be1 is surrounded by four H1 atoms to build the
tetrahedral structure and each Be2 connects with two H1 atoms and
two H2 atoms. This can be clearly seen from Fig. 1, in which the
larger blue spheres stand for Be1 atoms and the larger light blue
Be2 while the smaller gray spheres denote H1 atoms and the smaller
white H2. Calculated atomic structural parameters for orthorhombic
BeH$_{2}$ are presented in Table I. As a comparison, the previous
theoretical results \cite{Vajeeston, Hantsch} and available
experimental data \cite{Smith, Ahart} are also listed. It is clear
that our calculated results are on the whole in good agreement with
the experimental data, which thus confirms that the present
calculation is reliable and accurate. The exception is that in
analogy with previous theoretical report \cite{Hantsch}, our
calculated H1-Be2-H1 bond angle (98.3$^{\circ}$) is underestimated
by 12\% compared with the experimental value (112.1$^{\circ}$)
\cite{Smith}. Although it remains unclear for us to explain this
discrepancy, we can be sure that it is not caused by the choice of
the exchange-correlation potential. In fact, we have found that the
use of local density approximation (LDA) even more underestimates
the H1-Be2-H1 bond angle.

\begin{table}[ptb]
\caption{Calculated elastic constant of BeH$_{2}$ (\emph{Ibam}). All
values
are in units of GPa.}%
\label{tab:table2}%
\begin{ruledtabular}
\begin{tabular}{cccccccccccc}
\emph{C$_{11}$}&\emph{C$_{22}$}&\emph{C$_{33}$}&\emph{C$_{44}$}&\emph{C$_{55}$}&\emph{C$_{66}$}&\emph{C$_{12}$}&\emph{C$_{23}$}&\emph{C$_{13}$}\\
\hline
63.85 & 40.33  & 89.44 & 14.69  & 34.00 &21.01&0.63&0.66&10.76\\
\end{tabular}
\end{ruledtabular}
\end{table}

Elastic constants can measure the resistance and mechanical features
of crystal to external stress or pressure, thus describing the
stability of crystals. For the present orthorhombic BeH$_{2}$, there
are nine independent elastic constants, which are in our
first-principles calculations obtained by applying small elastic
strains. For each time that the strain changes, the coordinates of
ions are fully relaxed to ensure the energy minimum rule. In our
calculations the strain $\delta$ is varied in steps of 0.006 from
$\delta $=-0.036 to 0.036. The calculated elastic constants are
listed in Table II. Although at present there exist no experimental
data to compare with our calculated elastic constants, based on the
good agreement between our calculations and the experiment for the
atomic structural parameters, we expect that our calculated elastic
constants can be adopted in the realistic application if the
mechanical data are needed for BeH$_{2}$.

The availability of the elastic constants can be used to measure the stability
of a specific material. For the orthorhombic crystalline structure, the
mechanical stability criteria are given by \cite{Nye2}
\begin{align}
&  C_{11}>0,C_{22}>0,C_{33}>0,C_{44}>0,C_{55}>0,C_{66}>0,\nonumber\\
&  [C_{11}+C_{22}+C_{33}+2(C_{12}+C_{13}+C_{23})]>0,\nonumber\\
&  (C_{11}+C_{22}-2C_{12})>0,(C_{11}+C_{33}-2C_{13})>0,\nonumber\\
&  (C_{22}+C_{33}-2C_{23})>0.
\end{align}
From Table II we can see that the orthorhombic BeH$_{2}$ is mechanically
stable because its elastic constants satisfy formula (1). In fact, the
calculated amplitudes of \emph{C$_{12}$} and \emph{C$_{23}$} are negligibly
small, and the amplitude of \emph{C$_{13}$} is prominently smaller than that
of either \emph{C$_{11}$} or \emph{C$_{33}$}. Thus, formula (1) is easily satisfied.

After obtaining elastic constants, we can calculate bulk and shear moduli from
the Voigt-Reuss-Hill (VRH) approximation \cite{Voigt,Reuss,Hill}. The Voigt
bounds \cite{Voigt,Watt} on the effective bulk modulus \emph{B$_{V}$} and
shear modulus \emph{G$_{V}$} are
\begin{equation}
B_{V}=(1/9)[C_{11}+C_{22}+C_{33}+2(C_{12}+C_{13}+C_{23})]
\end{equation}
and
\begin{align}
G_{V}  &  =(1/15)[C_{11}+C_{22}+C_{33}+3(C_{44}+C_{55}+C_{66})\nonumber\\
&  -(C_{12}+C_{13}+C_{23})].
\end{align}
Under Reuss approximation \cite{Reuss}, the Reuss bulk modulus
\emph{B$_{R}$} and Reuss shear modulus \emph{G$_{R}$} are
\begin{align}
B_{R}  &  =\Delta\lbrack C_{11}(C_{22}+C_{33}-2C_{23})+C_{22}(C_{33}%
-2C_{13})\nonumber\\
&  -2C_{33}C_{12}+C_{12}(2C_{23}-C_{12})+C_{13}(2C_{12}-C_{13})\nonumber\\
&  +C_{23}(2C_{13}-C_{23})]^{-1}%
\end{align}
and
\begin{align}
G_{R}  &  =15\{4[C_{11}(C_{22}+C_{33}+C_{23})+C_{22}(C_{33}+C_{13})\nonumber\\
&  +C_{33}C_{12}-C_{12}(C_{23}+C_{12})-C_{13}(C_{12}+C_{13})\nonumber\\
&  -C_{23}(C_{13}+C_{23})]/\Delta+3[(1/C_{44})+(1/C_{55})\nonumber\\
&  +(1/C_{66})]\}^{-1},
\end{align}
where
\begin{align}
\Delta &  =C_{13}(C_{12}C_{23}-C_{13}C_{22})+C_{23}(C_{12}C_{13}\nonumber\\
&  -C_{23}C_{11})+C_{33}(C_{11}C_{22}-C_{12}^{2}).
\end{align}
The bulk modulus \emph{B} and shear modulus \emph{G}, based on Hill
approximation \cite{Hill}, are arithmetic average of Voigt and Reuss elastic
modulus, i.e., \emph{B}=$\frac{1}{2}(B_{R}+B_{V})$ and \emph{G}=$\frac{1}%
{2}(G_{R}+G_{V})$. The Young's modulus \emph{E} and Poisson's ratio $\upsilon$
for an isotropic material are given by \cite{Ravindran}
\[
E=\frac{9BG}{3B+G}%
\]
and
\[
\upsilon=\frac{3B-2G}{2(3B+G)}.
\]

The calculated results for these moduli and Poisson's ratio are
listed in Table I. Note that we have also calculated the bulk
modulus \emph{B} by fitting the Murnaghan equation of state. The
derived bulk modulus is well comparable with that from the above VRH
approximation, which again indicates that our calculations are
consistent and reliable. We find that the bulk modulus of BeH$_{2}$
is much smaller than that of MgH$_{2}$ (previous first-principles
reports \cite{Rici,Pfrommer} gave \emph{B}$\sim$50 GPa in the
MgH$_{2}$ system), which means that BeH$_{2}$ is more easily
compressed. The reason is that the rutile structure of MgH$_{2}$ is
more dense in atomic packing than the body-centered orthorhombic
structure of BeH$_{2}$. Although at present no experimental data for
these moduli and Poisson's ratio are attainable for pure BeH$_{2}$,
recent Brillouin scattering experiment \cite{Ahart} on amorphous
BeH$_{2}$ gave \emph{B}=14.2$\pm$3.0 GPa and $\upsilon$=0.21.
Considering the fact that the amorphous BeH$_{2}$ is more loose than
the pure BeH$_{2}$, it is understandable that our calculated modulus
for the pure phase is larger than the experimental measurement for
the amorphous phase. Concerning the Poisson's ratio, our calculated
$\upsilon $=0.102 for the pure BeH$_{2}$ is much smaller than the
experimental measurement on the amorphous phase. This is also really
understandable. In fact, it is well known that for the common
materials that have much smaller shear moduli compared with the bulk
moduli, their Poisson's ratio is close to 1/3. If the shear modulus,
on the other hand, is much larger than the bulk modulus, materials
will have negative Poisson's ratio \cite{Lakes}. In the case of
BeH$_{2}$, our calculated results in Table I show that the shear
modulus is comparable to the bulk modulus. Then, it is anticipated
that the Poisson's ratio of the orthorhombic BeH$_{2}$ has a small
value, which is consistent with our calculated result of
$\upsilon$=0.102. The experimentally measured larger Poisson's ratio
\cite{Ahart} is therefore due to the use of amorphous sample.

\subsection{Electronic structure and charge distribution}

\begin{figure}[tbp]
\begin{center}
\includegraphics[width=0.8\linewidth]{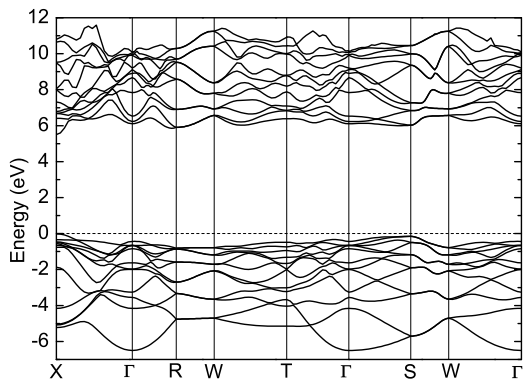}
\end{center}
\caption{GGA band structure of BeH$_{2}$. The Fermi energy is set at
zero.}
\end{figure}%

The calculated band structure of BeH$_{2}$ is shown in Fig. 2. Our
theoretical valence-band width is 6.5 eV and band gap 5.5 eV, which
is consistent with previous calculation \cite{Vajeeston}. As an
illustrating comparison, in isoelectronic MgH$_{2}$ the calculated
valence-band width is 7.5 eV and band gap 4.2 eV\cite{VajeestonPRL},
which implicitly suggests that the Mg$-$H bond is more ionic than
the Be$-$H bond. Also one can see from Fig. 2 that the valence band
maximum (VBM) and conduction band minimum (CBM) both appear at
\textbf{X} point in the Brillouin zone. By calculating the
orbital-decomposed band charges inside the muffin-tin spheres of
BeH$_{2}$ at VBM and CBM at \textbf{X} point, we find that the VBM
has predominant H $1s$ state character mixed with significant Be
$2p$ contribution, while the CBM has obvious Be $2s$ and $2p$
features mixed with a little H $1s$ contribution.

\begin{figure}[tbp]
\begin{center}
\includegraphics[width=1.0\linewidth]{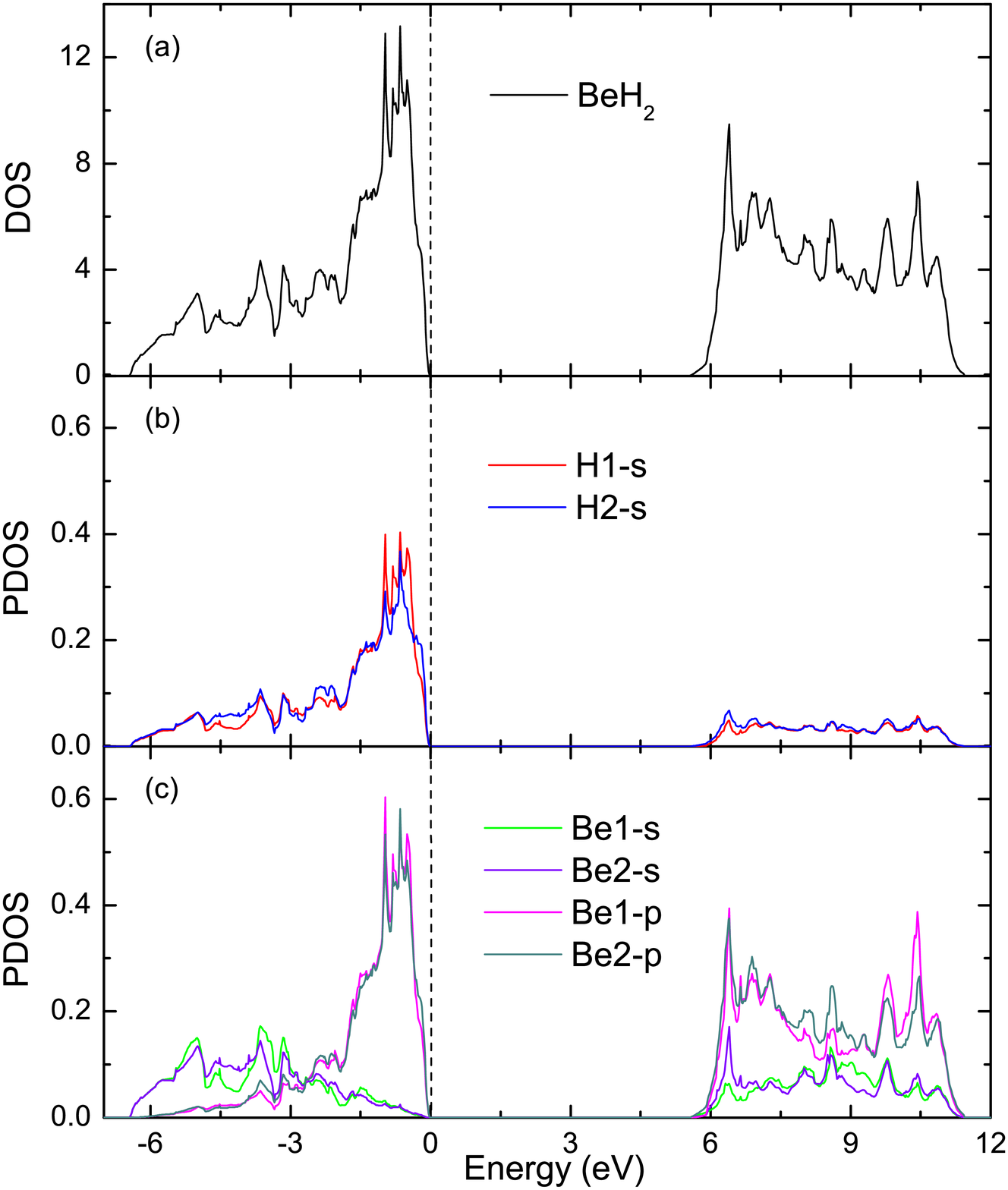}
\end{center}
\caption{(Color online) Total and orbital-resolved local densities
of states for ground state BeH$_{2}$. The Fermi energy is set at
zero.}
\end{figure}%

The total electronic density of states (DOS) per unit cell for the
orthorhombic BeH$_{2}$ is shown in Fig. 3(a). For more clear
illustration, the orbital-resolved partial DOS per atom in the unit
cell are also shown in Fig. 3, see Fig. 3(b) for H1 and H2 atoms,
and Fig. 3(c) for Be1 and Be2 atoms. From Fig. 3 the following
prominent features can be seen: (i) H $1s$ state hybridizes with Be
$2s$ and $2p$ states in the whole range of the valence band. This is
different from alkali-metal hydrides. Due to the combined fact that
there has only one valence electron with low ionization energy and
the $s{\small \rightarrow p}$ promotion energy is relatively large
in alkali-metal elements, usually alkali-metal hydrides have entire
ionic bonding in the sense that the valence electrons in
alkali-metal atoms totally transfer to the H atoms. This clear ionic
picture in alkali-metal hydrides becomes blurred in
alkaline-earth-metal hydrides. In the present case of BeH$_{2}$, in
addition to a large ionic weight (see the following
charge-distribution analysis), a distinct covalent component also
exists in the Be$-$H bond, which is responsible for the
stabilization of the low symmetric structure of BeH$_{2}$.
Prominently, this covalency is dominated by the hybridization of H
$1s$ and Be $2p$ orbitals near the top of the valence band, while
the hybridization of H $1s$ and Be $2s$ orbitals is relatively weak,
as shown in Fig. 3. Thus, the ionicity in the Be$-$H bond is mainly
featured by charge transfer from Be $2s$ to H $1s$ atomic orbitals.
Interestingly, in beryllium oxide the Be$-$O bond also exhibits
strong covalency. As a result, bulk BeO has wurtzite structure
instead of the usual form of rocksalt structure taken by other
alkaline-earth-metal oxides; (ii) In the range of conduction band
the DOS is mainly featured by Be $2s$ and $2p$ states, mixed with a
little H $1s$ states; (iii) Although there are three Be$-$H bond
configurations (Be1$-$H1, B1$-$H2, and Be2$-$H2), the DOS propertie
of these chemical bonds are almost the same and there is no
remarkable difference between Be1 and Be2 (or between H1 and H2)
atoms.

In order to gain more insight into the nature of Be$-$H chemical
bonding in the orthorhombic BeH$_{2}$, we have investigated the
valence charge density distribution. Two contour plots of the
valence charge density distributions are shown in Fig. 4. Here the
contour plane in Fig. 4(a) is the plane composed by the three
labeled atoms (1 to 3) shown in Fig. 1, while the contour plane in
Fig. 4(b) is the (001) plane (perpendicular to the \emph{c} axis).
It is these two planes that characterize the bond structure of
BeH$_{2}$. In both planes, hydrogen charge density shape is deformed
toward the direction to the nearest-neighbor Be atoms. The charge
density around H atoms is higher than that around Be atoms. This
further indicates aforementioned significant ionic-type charge
transfer from Be 2\emph{s} to H 1\emph{s} state. On the other hand,
the charge densities at the midpoint of Be$-$H bond is 0.46 e/{\AA
}$^{3}${, }which is prominently higher than that in typical ionic
crystals. This values is also much larger than that in isoelectroic
MgH$_{2}$ system. In fact, our calculation on MgH$_{2}$ gave charge
density of 0.17 e/{\AA }$^{3}$ at the midpoint of the Mg$-$H bond,
which is comparable well with recent experimental data
\cite{Noritake} of 0.2 e/{\AA }$^{3}$. Thus again, the charge
distribution reveals that hydrogen is more strongly bonded to Be in
BeH$_{2}$ than to Mg in MgH$_{2}$. As a result, the calculated
Be$-$H bond length ($\sim$1.37 {\AA , see Table I}) is remarkably
smaller than the Mg$-$H bond length (1.93 {\AA }). Although hydrogen
atoms do not form bond, for illustration, here we also present the
charge densities at the midpoint in the line that links the
nearest-neighbor H atoms, which are $\sim$0.2 e/{\AA }$^{3}$ for
BeH$_{2}$ and $\sim$0.15 e/{\AA }$^{3}$ for MgH$_{2}$.
Interestingly, the nearest-neighbor H-H distance in BeH$_{2}$ is
also larger than that in MgH$_{2}$ by $\sim$0.25 {\AA .}

\begin{figure}[tbp]
\begin{center}
\includegraphics[width=1.0\linewidth]{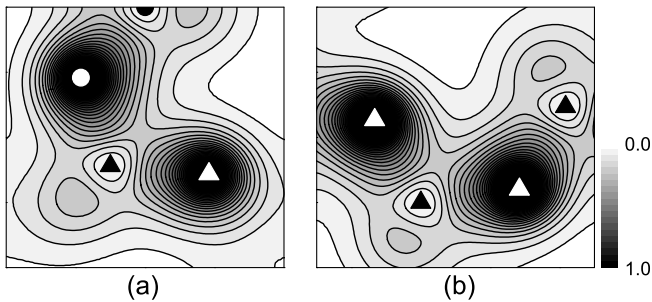}
\end{center}
\caption{Valence charge density maps of BeH$_{2}$ where (a) is the
plane established by the three labeled atoms (1 to 3) as shown in
Fig. 1 and (b) is the (001) plane. Here $\bullet$ stands for Be1,
$\blacktriangle$ for Be2, $\circ$ for H1 and $\vartriangle$ for H2.
The contour lines are drawn from 0.0 to 1.0 at 0.05 ev/{\AA}$^{3}$
intervals.}
\end{figure}%

From above results, now it becomes clear that the bonding nature of
hydrogen in BeH$_{2}$, including the ionicity and covalency, is even
more complex than in MgH$_{2}$. Specially, because of complicated
charge transfer process, which involves not only ionic but also
covalent charge redistribution, one may wonder how to characterize
the valency of H and Be in BeH$_{2}$. Without knowing more
appropriate way to valency, here we define and determine the
so-called ionic radii of H and Be by using the method raised by Yu
and Lam \cite{Rici} in studying MgH$_{2}$. In this way, H ionic
radius is defined as the half of average nearest-neighbor H-H
distance. Subtracting the H ionic radius from the average
nearest-neighbor Be-H distance gives the ionic radius of Be. Thus
the ionic radius of H in our system is 1.131 $\mathrm{{\mathring
{A}}}$ and 0.237 {\AA } for Be ions. After integration, we find
0.005 electrons around Be and 1.631 electrons around H. As a result,
the valency of Be and H are represented as Be$^{1.99+}$ and
H$^{0.63-}$. For MgH$_{2}$ our calculations give the valency of Mg
and H as Mg$^{1.95+}$H$^{0.60-}$. Clearly, our calculated valency
shows that compared to the MgH$_{2}$ system, charge transfer process
is more prominent in BeH$_{2}$ system. Therefore, we would like to
derive that the the definition of ionic radius by Yu and Lam still
works for more covalent BeH$_{2}$ hydride. \\
\section{CONCLUSION}

In summary, we have investigated the structural, mechanical, electronic, and
chemical bonding properties of BeH$_{2}$ through first-principles DFT-GGA
calculations. The calculated structural parameters are in good agreement with
available experimental results. We have calculated the nine independent
elastic constants and our results well obey the mechanical stability criteria.
In particular, the calculated \emph{C$_{12}$} and \emph{C$_{23}$} are
negligibly small, and \emph{C$_{13}$} is much smaller than \emph{C$_{11}$} or
\emph{C$_{33}$}. The bulk modulus \emph{B}, shear modulus \emph{G}, Young's
modulus \emph{E} and Poisson's ratio $\upsilon$ have been obtained based on
the knowledge of the elastic constants. The bonding nature of hydrogen in
BeH$_{2}$ have been fully analyzed in terms of the band structure, density of
state, and valence charge distribution. It has been shown that the Be$-$H bond
displays a mixed ionic/covalent character. Here the ionicity is mainly
featured by charge transfer from Be 2\emph{s} to H 1\emph{s} states, while the
covalency is manifested by prominent hybridization of H 1\emph{s} and Be
2\emph{p} states. The elemental valency in BeH$_{2}$ has been determined to be
Be$^{1.99+}$ and H$^{0.63-}$. By a systematic comparison, it has been shown
that the chemical bonding nature of hydrogen in BeH$_{2}$ is even more complex
than in MgH$_{2}$.


\begin{thebibliography}{99}                                                                                               %


\bibitem {Armstrong}D. A. Armstrong, J. Jamieson, and P. G. Perkins, Theor.
Chim. Acta. \textbf{51}, 163 (1979).

\bibitem {Smith}G. S. Smith, Q. C. Johnson , D. K. Smith, D. E. Cox, R. L.
Snyder, and R. S. Zhou, Solid. State. Commun. \textbf{67}, 491 (1988).

\bibitem {Vajeeston}P. Vajeeston, P. Ravindran, A. Kjekshus, and H.
Fjellv{\aa }g, Appl. Phys. Lett. \textbf{84}, 34 (2004).

\bibitem {Hantsch}U. Hantsch, B. Winkler, and V. Milman, Chen. Phys. Lett.
\textbf{378}, 343 (2003).

\bibitem {Ahart}M. Ahart, J. L. Yarger, K. M. Lantzky, S. Nakano, H. Mao, and
R. J. Hemley, J. Chem. Phys. \textbf{124}, 014502 (2006).

\bibitem {Kresse3}G. Kresse and J. Furthm\"{u}ller, Phys. Rev. B \textbf{54},
11169 (1996).

\bibitem {PAW}P. E. Bl\"{o}chl, Phys. Rev. B \textbf{50}, 17953 (1994).

\bibitem {GGA}J. P. Perdew, K. Burke, and Y. Wang, Phys. Rev. B \textbf{54},
16533 (1996).

\bibitem {Monk}H. J. Monkhorst and J. D. Pack, Phys. Rev. B \textbf{13}, 5188 (1972).

\bibitem {Nye2}J. F. Nye, \textit{Physical Properties of Crystals} (Oxford
University Press, 1985).

\bibitem {Voigt}W. Voigt, \textit{Lehrburch der Kristallphysik} (Teubner,
Leipzig, 1928).

\bibitem {Reuss}A. Reuss and Z. Angew, Math. Mech. \textbf{9}, 49 (1929).

\bibitem {Hill}R. Hill, Phys. Soc. London \textbf{65}, 350 (1952).

\bibitem {Watt}J. P. Watt, J. Appl. Phys. \textbf{50}, 6290 (1979).

\bibitem {Ravindran}P. Ravindran, L. Fast, P. A. Korzhavyi, B. Johansson, J.
Wills, and O. Eriksson, J. Appl. Phys. \textbf{84}, 4891 (1998).

\bibitem {Lakes}R. S. Lakes, Nature (London) \textbf{358}, 713 (1992).

\bibitem {VajeestonPRL}P. Vajeeston, P. Ravindran, A. Kjekshus, and H.
Fjellv{\aa }g, Phys. Rev. Lett. \textbf{89}, 175506 (2002).

\bibitem {Noritake}T. Noritake, M. Aoki, S. Towata, Y. Seno, Y. Hirose, E.
Nishibori, M. Takata, and M. Sakata, Appl. Phys. Lett. \textbf{81}, 2008 (2002).

\bibitem {Rici}R. Yu and P. K. Lam, Phys. Rev. B \textbf{37}, 8730 (1988).

\bibitem {Pfrommer}B. Pfrommer, C. Els\"{a}sser, and M. F\"{a}hnle Phys. Rev.
B \textbf{50}, 5089 (1994).

\end{thebibliography}
\end{document}